\documentclass[10pt,aps,prl,groupedaddress,showpacs,twocolumn,showkeys]{revtex4-1}
\usepackage{amsmath}
\usepackage{amssymb}
\usepackage{amsfonts}
\usepackage{bm}
\usepackage{graphicx}
\usepackage{epstopdf}
\usepackage{float}
\usepackage{caption}
\usepackage{subcaption}
\usepackage{comment}
\usepackage{mathrsfs}
\usepackage[USenglish,american]{babel}
\usepackage[OT2,T1]{fontenc}
\DeclareSymbolFont{cyrletters}{OT2}{wncyr}{m}{n}
\DeclareMathSymbol{\Sha}{\mathalpha}{cyrletters}{"58}
\usepackage{color}

\begin{document}
\title{Modal Expansion of the Impulse Response Function for Predicting the
Time Dynamics of Open Optical Cavities}
\author{R\'{e}mi Colom$^{1,2}$, Brian Stout$^3$ and Nicolas Bonod$^3$}

\address{$^1$ Avignon Universit\'{e}, UMR 1114 EMMAH, Avignon Cedex 84018, France}
\address{$^2$ Zuse Institute Berlin, Takustra{\ss}e 7, 14195 Berlin, Germany}
\address{$^3$ Aix Marseille Univ, CNRS, Centrale Marseille, Institut Fresnel, 13013 Marseille, France}

\begin{abstract}
Light interaction with optical cavities is of fundamental interest to enhance the light-matter interaction and to shape the spectral features of the electromagnetic fields. Important efforts have been carried out to develop modal theories of open optical cavities relying on an expansion of the fields on the eigen-fields of the cavity. Here, we show how such an expansion predicts the temporal dynamics of optical resonators. We consider a Fabry-Perot cavity to derive the full analytical expressions of the internal and scattered field on the quasi-normal modes basis together with the complex eigen-frequencies. We evince the convergence and accuracy of this expansion before deriving the impulse response function (IRF) of the open cavity. We benefit from this modal expansion and IRF to demonstrate that the eigen-modes of the open cavity impact the signals only during the transient regimes and not in the permanent regime.
\end{abstract}

\maketitle

\section{Introduction}
Resonances play a fundamental role in various fields of wave physics, such as acoustics, electronics, mechanics and electromagnetism. In optics and photonics, resonant light interaction with optical resonators is a key concept that plays a role in a variety of applications such as laser cavities \cite{sandoghdar1996very}, non linear optics \cite{del2007optical} and enhanced Purcell factors \cite{somaschi2016near}. The simplest and most widespread type of cavity is certainly the slab geometry, also known as the Fabry-Perot cavity \cite{Born00,baranov2017coherent}. 

Resonances originate from the coupling between the eigen-modes of the photonic cavities with the excitation field \cite{lalanne2018light,Kristensen2019,krasnok2019anomalies}. Optical cavities are never perfect and light must eventually either leak outside the cavity or be absorbed by material degrees of freedom. Consequently, optical cavities are mathematically described by non-Hermitian operators and their eigen-frequencies are complex $\omega_n = \omega'_n + i \omega''_n$. Eigen-modes in non-Hermitian systems are called Quasi-Normal Modes (QNMs) or resonant states~\cite{lalanne2018light,Kristensen2019}. Causality imposes that $\omega''_n < 0$ when the $e^{-i\omega t}$ convention for time dependence is used. Scattering losses are described by imposing that the QNM of these structures satisfy outgoing boundary conditions. They consequently behave as $e^{i k_n r}$ with $k_n = \frac{\omega_n}{c}$ when $r\rightarrow\infty$ which yield an exponential divergence from the photonic structure also called ``exponential catastrophe''~\cite{Thomson1883,Lamb1900,nussenzveig1972causality}. This problem, and the related issue of the normalization of the modes, can be addressed in the harmonic domain with the use of Perfectly Matched Layers~\cite{Sauvan2013,faggiani2017modal,lalanne2019quasinormal,Yan2018} or of Gaussian regularization~\cite{stout2019eigenstate}. The divergence may also be tackled in the time domain when causality is enforced~\cite{Colom2018,abdelrahman2018completeness}. Expansion of the internal field may however still be carried out despite the divergence~\cite{Zschiedrich2018,muljarov2011brillouin,Leung1996}. It turns out that the QNM theory is particularly well suited to study resonant light interaction with open optical cavities in the time domain. In particular, this theory is expected to provide novel insights in the dynamics of the optical response of photonic systems. 

Here we provide the modal and analytical expressions of the reflected, transmitted and internal fields of a Fabry-Perot cavity in the harmonic and time domains. A comparison between analytic and numerical calculations of the reflected and transmitted fields shows the convergence and strong accuracy of these analytical expressions. We therefore show how to derive the impulse response function of the optical system from which can be obtained any response from an arbitrary excitation. We benefit from these expressions to demonstrate that the eigen-modes of the cavity excited by a sinusoidal causal field significantly contribute to the optical response during the transient regime but not in the permanent regime. We derive fully analytical expressions for both eigen-frequencies and fields. This work can therefore serve as a benchmark for other numerical methods. Let us emphasize that the $S$ and $T$-matrix formalism can be applied to a wide range of structures such as diffraction gratings and even arbitrarily shaped scatterers through a multipolar decomposition~\cite{Gippius2010,grigoriev2013singular,Weiss2018,demesy2018scattering}. The method derived in this work is therefore general and can be applied to other geometries of photonic cavities. 

\section{Modal expansion of the optical response in the time domain}
Let us consider an incident field which is a linearly polarized plane wave at normal incidence on a Fabry-Perot cavity of refractive index $n_s$ and thickness $d$ placed in air \cite{baranov2017coherent}. In this configuration, the response of the system is polarization independent and the field can be described by a scalar model: $E_{\text{inc}}(z,t)=E(\omega)e^{ikz}e^{-i\omega t}$.
The expression of the reflected $\mathbf{E}_{\tilde{r}}(z,t)$ or transmitted $\mathbf{E}_{\tilde{t}}(z,t)$ fields with respect to time $t$ and ordinate $z$ is: $\mathbf{E}_{\tilde{r},\tilde{t}}(z,t)=\int_{-\infty}^{\infty} (\tilde{r}(\omega),\tilde{t}(\omega)) E_{\text{inc}}(z=0,\omega) e^{-ikz} e^{-i \omega t} d\omega$
The reflection and transmission coefficients can be derived from the $S$-matrix diagonalized with even $(e)$ and odd $(o)$ modes (see Supplemental Material): 
$\tilde{r}(\omega) = \tilde{r}_{\text{n.r.}}-\frac{1}{2}\sum_{\alpha=-M}^{M}\frac{r_{\alpha}^{(e)}}{\omega-\omega_{p,\alpha}^{(e)}}-\frac{1}{2}\sum_{\alpha=-M}^{M}\frac{r_{\alpha}^{(o)}}{\omega-\omega_{p,\alpha}^{(o)}}$ and $\tilde{t}(\omega) = -\frac{1}{2}\sum_{\alpha=-M}^{M}\frac{r_{\alpha}^{(e)}}{\omega-\omega_{p,\alpha}^{(e)}}+\frac{1}{2}\sum_{\alpha=-M}^{M}\frac{r_{\alpha}^{(o)}}{\omega-\omega_{p,\alpha}^{(o)}}$, where $\tilde{r}_{\text{n.r.}} = -\frac{\sum_{\alpha=-M}^{M}\frac{r_{\alpha}^{(e)}}{p_{\alpha}^{(e)}}+\sum_{\alpha=-M}^{M}\frac{r_{\alpha}^{(o)}}{p_{\alpha}^{(o)}}}{2}$. The eigen-frequencies can be computed analytically in the case of Fabry-Perot cavities and correspond respectively to the even and odd elements of this sequence: 
$\omega_{p,\alpha}=\frac{\pi \alpha c}{nd}+i \frac{\ln(r')c}{nd}$. The residues possess the following expression $r_{\alpha,\omega}^{(e)} = r_{\alpha,\omega}^{(o)} = \frac{1-\left(r'\right)^2}{r' i \frac{d}{c} n}$ and are thus independent of the eigen-frequency and the type ({\it i.e} even or odd) of the associated mode. The use of the pole expansion of $\tilde{r}$ and $\tilde{t}$ along with the residue theorem and the convolution theorem allow us to derive the following expressions for the time-dependent reflected and transmitted fields:
\begin{widetext}
\begin{eqnarray}
\begin{aligned}
&E_{\tilde{r}}(z,t)=E_{\text{inc}}(z = 0,t) \ast \left[\tilde{r}_{\rm  n.r.}\delta\left(t+\frac{z}{c}\right)+i\frac{H\left(t+\frac{z}{c}\right)}{2}\sum_{\alpha=-M}^{M}\left(r_{\omega,\alpha}^{(e)}e^{-i \omega_{p,\alpha}^{(e)} \left(t+\frac{z}{c}\right)}+r_{\omega,\alpha}^{(o)}e^{-i \omega_{p,\alpha}^{(o)} \left(t+\frac{z}{c}\right)}\right)\right] \\
&\\
&E_{\tilde{t}}(z,t) =  E_{\text{inc}}(z=0,t) \ast \left[\frac{i}{2}H\left(t-\frac{z}{c}\right)\right.\left.\left(\sum_{\alpha=-M}^{M} r_{\omega,\alpha}^{(e)} e^{-i \omega_{p,\alpha}^{(e)} \left(t-\frac{z}{c}\right)}-r_{\omega,\alpha}^{(o)}e^{-i \omega_{p,\alpha}^{(o)} \left(t-\frac{z}{c}\right)}\right)\right] \\
\end{aligned} 
\label{E_R_T_express}
\end{eqnarray}
\end{widetext}
Expression of the internal is provided in Eq.~(33) of the Supplemental Material. Eqs.~(\ref{E_R_T_express}) show that the reflected and transmitted fields can be expanded on the QNM functions proportional to $e^{-i \omega_{p,\alpha}^{(i)} \left(t-\frac{z}{c}\right)}$. The basis has to be numerically truncated to $2M+1$ modes. In such
a configuration, we can calculate the scattered fields with a quasi-normal mode (QNM) expansion without the need of any fitting parameter or numerical approach. Let us stress that despite the presence of exponentially diverging terms $e^{-i \omega_{p,\alpha}^{(i)} \left(t-\frac{z}{c}\right)}$ with $\Im(\omega_{p,\alpha}^{(i)})<0$, the previous expansions do not diverge thanks to the presence of the Heaviside distribution in the expression $H\left( t-\frac{z}{c} \right)$ \cite{Colom2018}. The main difference between the reflected and transmitted/internal fields is the presence of a term $\tilde{r}_{\rm n.r.}$ in the reflected field \cite{Colom2018,colom2019modal}. 

\begin{widetext}
\begin{center}
\begin{figure}[h]
\includegraphics[width=\textwidth]{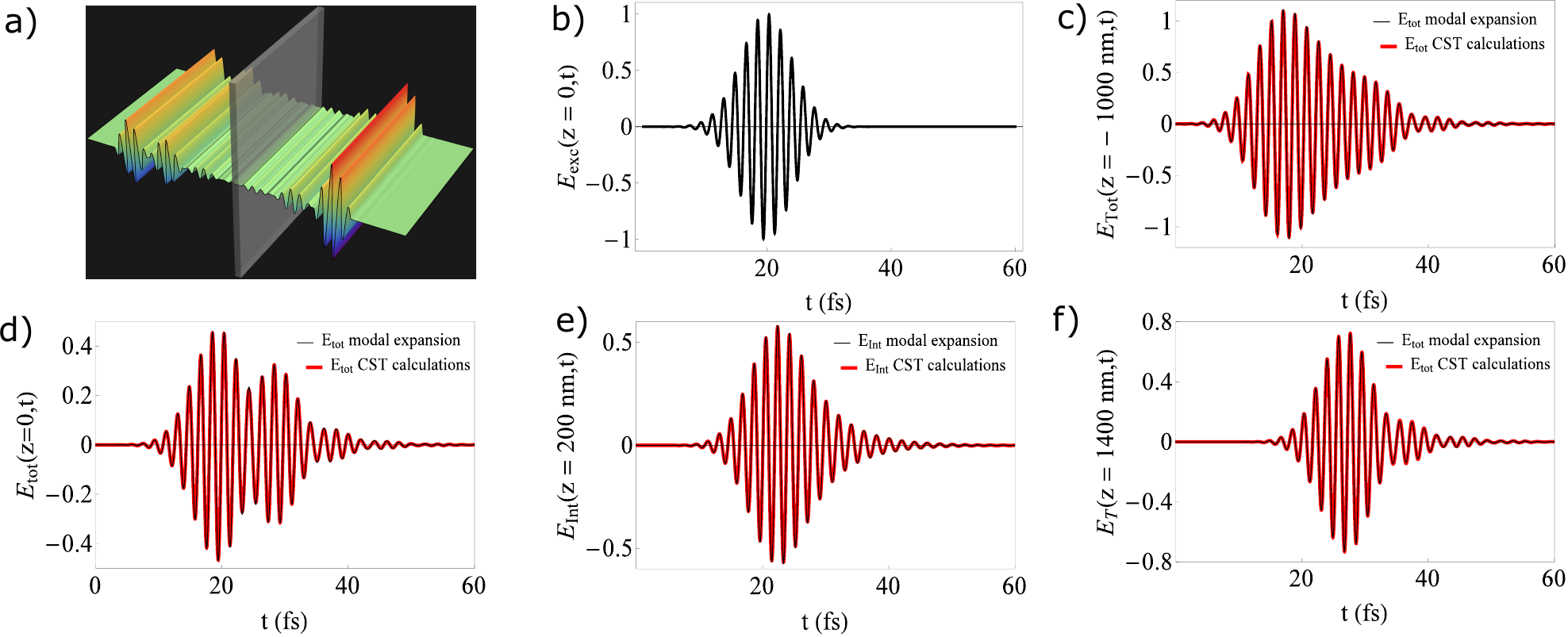}
\caption{Optical response of a slab in the time domain. a) Simulation setup: an exciting field propagating along the $z$-axis with a pulse with a Gaussian envelope impinges at normal incidence on a slab of thickness $d=400$~nm, composed of a material of refractive index $n_s=3$. The ordinate $z=0$ is taken at the left interface. b). Amplitude of the incident electric field at $z=0$ with respect to time (in sec). (c-f). Total (incident+reflected) electric field amplitude calculated on the incident medium ($z=-1000$~nm (c), and $z=0$~nm (d)), internal electric field calculated in the middle of the slab ($z=200$~nm (e)), and transmitted electric field calculated 1000~nm after the slab ($z=1400$~nm (f)). The electric field amplitude is plotted with respect to time and calculated with analytical expressions in Eqs.~\ref{E_R_T_express} and Eq.~(33) of the Supplemental Material (red lines) and with CST (black lines).}
\label{Fig:1} 
\end{figure}
\end{center}
\end{widetext}

The validity of this modal expansion is assessed by studying a vertical slab illuminated by a pulse (see Fig.~\ref{Fig:1}(a-b)). The slab is in air and is composed of a material of refractive index of $n_s=3$, its thickness is $400$~nm. We consider a sinusoidal electric field excitation with a Gaussian envelope $E_{\text{exc}}(z,t)= \Pi\left(\frac{t-\frac{z}{c}-\frac{z_s}{c}-\frac{a}{2}}{2}\right)e^{-\left(\frac{t-\frac{z}{c}-\frac{z_s}{c}-\frac{a}{2}}{\tau}\right)^2}
\sin\left(\omega\left(t-\frac{z}{c}-\frac{z_s}{c}-\frac{a}{2}\right)\right)$ where $z_s = 1000$ nm, $\omega = 2 \pi f$, $f = 545$~THz, $\tau=\frac{20}{2\pi f}$ and $a=\frac{36\pi}{\omega}$ (see Fig.~\ref{Fig:1}(a)). $\Pi$ is the door function that has been multiplied to the excitation field to clearly specify the time at which the excitation field reaches the slab. These calculations are performed by taking into account $M = 100$ terms in the modal expansions. It can be observed that the calculations performed with the analytical expressions in Eqs.~(\ref{E_R_T_express}) match very finely to those obtained with CST in the time domain. 

\section{Impulse response function of an optical resonator}
\begin{figure}[h!]
\begin{center}
\includegraphics[width=0.5\textwidth]{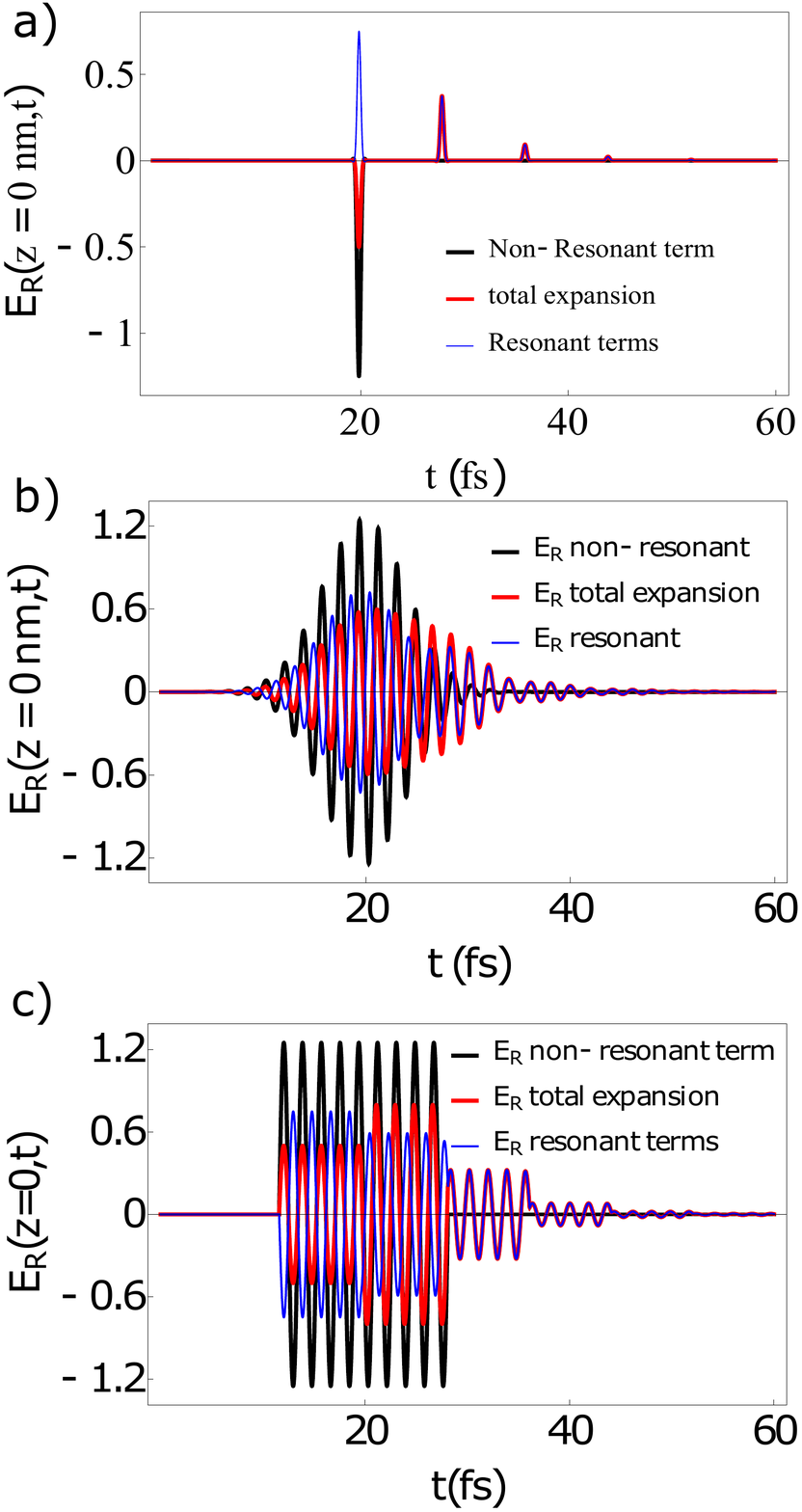}
\end{center}
\caption{Reflected electric field at $z=0$ for the different envelopes of the time-dependent excitation. In a) a very narrow Gaussian excitation is used to illustrate and the reflected field is close to the impulse response function. In b), the excitation has a Gaussian envelope. In c), the envelope of the excitation is a rectangular function.}
\label{IRF} 
\end{figure}

Impulse Response Function (IRF), which is related to the time-dependant Green function, is a key function to analyze systems in wave physics since it permits to retrieve the time-dependent response of the system excited by any excitation field.
The IRF corresponds to the case where $E_{\text{inc}}(z=0,t) = \delta(t)$. For $E_{\tilde{r}}(z,t)$ we obtain:
\begin{eqnarray}
\begin{aligned}
&E_{\tilde{r}}(z,t) = \tilde{r}_{\rm n.r.}\delta\left(t+\frac{z}{c}\right)+i\frac{H\left(t+\frac{z}{c}\right)}{2}\\
&\sum_{\alpha=-M}^{M}\left(r_{\omega,\alpha}^{(e)}e^{-i \omega_{p,\alpha}^{(e)} \left(t+\frac{z}{c}\right)}+r_{\omega,\alpha}^{(o)}e^{-i \omega_{p,\alpha}^{(o)} \left(t+\frac{z}{c}\right)}\right)
\end{aligned}
\label{E_R_express_Dirac}
\end{eqnarray}
Similar expressions are obtained for $E_{\tilde{t}}(z,t)$ and $E_{\text{int}}(z,t)$ (see Eq.~(35) of the Supplemental Material). This shows that the response of a Fabry-Perot cavity to a Dirac excitation is a superposition of one Dirac distribution and a set of oscillating sinusoidal functions with an exponentially decreasing envelope. However, one would expect that the response of a Fabry-Perot to a Dirac excitation would be a series of Dirac distributions with a decreasing envelope depending on the number of times light has been reflected back and forth inside the slab. Is this expected result in contradiction with the expression provided in Eq.~\ref{E_R_express_Dirac}? To answer this question, we first need to simplify this expression. For that purpose, let us point out that $\Im\left(\omega_{\alpha}^{(o)}\right)=\Im\left(\omega_{\alpha}^{(e)}\right) =\frac{\ln(r')c}{nd}$ which allows us to get the simplified expression of the reflected field:
\begin{eqnarray}
\begin{aligned}
&E_{\tilde{r}}(z,t) = \tilde{r}_{\rm n.r.}\delta\left(t+\frac{z}{c}\right)+i\frac{H\left(t+\frac{z}{c}\right)}{2}r_{\omega}e^{\frac{\ln(r')c}{nd}\left(t+\frac{z}{c}\right)}\\
&\sum_{\alpha=-M}^{M}\left(e^{-i \omega_{p,\alpha}^{(e)'} \left(t+\frac{z}{c}\right)}+e^{-i \omega_{p,\alpha}^{(o)'} \left(t+\frac{z}{c}\right)}\right)= R_{\rm n.r}\delta\left(t+\frac{z}{c}\right)\\
&+i\frac{H\left(t+\frac{z}{c}\right)}{2}r_{\omega}e^{\frac{\ln(r')c}{nd} \left(t+\frac{z}{c}\right)}\sum_{\alpha=-2M}^{2M}e^{-i 2\pi c\frac{\alpha}{2nd} \left(t+\frac{z}{c}\right)}.
\label{E_R_express_simpl}
\end{aligned}
\end{eqnarray}

where we used $\omega_{p,\alpha}^{(e)'}=\frac{2\pi \alpha c}{nd}$ and $\omega_{p,\alpha}^{(o)'}=\frac{\pi (2 \alpha-1) c}{nd}$, the Dirac comb being equal to $\Sha_{T}(t)=\sum_{\alpha=-\infty}^{\infty}\delta(t-\alpha T)=\frac{1}{T}\sum_{\alpha=-\infty}^{\infty}e^{-i\frac{2\pi\alpha}{T}t}$, the sum in Eq.~(\ref{E_R_express_simpl}) is consequently approximately equal to: $ \sum_{\alpha=-M}^{M}e^{-i 2\pi c\frac{\alpha}{2nd}\left(t+\frac{z}{c}\right)} \approx \frac{2nd}{c}\sum_{\alpha=-M}^{M}\delta(t+\frac{z}{c}-\alpha\frac{2nd}{c})$. 
It turns out that the expression in Eq.~(\ref{E_R_express_Dirac}) is approximately a superposition of Dirac distributions with exponentially decreasing amplitudes and a period equal to $\frac{2nd}{c}$. The resonant terms in Eqs.~(\ref{E_R_express_Dirac}) interfere to create a series of peaks with a decreasing amplitude. This result is consequently in agreement with physical expectations.
The impact of the truncation on the sum in Eq.~(\ref{E_R_express_simpl}) can be analyzed by studying the reflected field produced by a very short excitation field. In Fig.~\ref{IRF}, we show the reflected field at $z=0$ created when the dielectric slab is illuminated by an excitation field defined by $E_{\text{exc}}(z,t)= \Pi\left(\frac{t-\frac{z}{c}-\frac{z_s}{c}-\frac{a}{2}}{2}\right)e^{-\left(\frac{t-\frac{z}{c}-\frac{z_s}{c}-\frac{a}{2}}{\tau}\right)^2}\cos\left(\omega\left(t-\frac{z}{c}-\frac{z_s}{c}-\frac{a}{2}\right)\right)$, where $\tau=\frac{1}{2\pi f}$ while all the other parameters are the same as in the previous section. 

This example illustrates the fact that the reflected field is a series of peaks with a decreasing amplitude. It helps us to clarify the role of each term in Eq.~(\ref{E_R_express_simpl}). It is in particular noticeable that there is a huge difference in the first peak when the non-resonant term is taken into account. This first peak corresponds to the direct reflection of light at the first interface of the dielectric slab. One consequently expects its amplitude to be equal to $r$, the Fresnel coefficient since the amplitude of the excitation field is equal to 1. In this case the Fresnel coefficient is equal to $r = \frac{1-3}{3+1} = -0.5$ since the refractive index is $n_s = 3$. This is indeed what is observed when all the terms are taken into account while the amplitude of the first peak is far from this value when only resonant terms are taken into account. This unveils the role of the non-resonant term that guarantees the boundary conditions at the interface. At larger time, $i.e.$ when the excitation is over, the response is fully described by the resonant terms. Figs.~\ref{IRF} b and c) illustrate the versatility of this method as it allows to compute the response from any type of excitation.

\section{From transient to permanent regime}
Let us now establish the role of QNMs fields in transient and permanent regimes. Let us calculate the optical response of a Fabry-Perot resonator for an excitation equal to $E_{\text{inc}}(z=0,t) =H\left(t\right)\sin\left(\omega_0 t\right)=\Re\left(iH\left(t\right)e^{-i \omega_0 t}\right)$. Derivations are carried out in Supplemental Material and lead to the following expression for the reflected field:

\begin{eqnarray}
\begin{aligned}
&E_{\tilde{r}}(z,t)= \Re\left\{ \tilde{r}(\omega_0)iH\left(t+\frac{z}{c}\right)e^{-i \omega_0 \left(t+\frac{z}{c}\right)}\right.\\
&\left.+\frac{iH\left(t+\frac{z}{c}\right)}{2}\sum_{\alpha=-M}^{M}\left[\frac{r_{\omega,\alpha}^{(e)}}{\omega_{0}-\omega_{p,\alpha}^{(e)}}  e^{-i \omega_{p,\alpha}^{(e)} \left(t+\frac{z}{c}\right)}\right.\right. \\
&\left.\left.+ \frac{r_{\omega,\alpha}^{(o)}}{\omega_{0}-\omega_{p,\alpha}^{(o)}} e^{-i \omega_{p,\alpha}^{(o)} \left(t+\frac{z}{c}\right)}\right]\right\}
\end{aligned}
\label{E_R_express_bis}
\end{eqnarray}

\begin{figure}[h!]
\begin{center}
\includegraphics[width=0.5\textwidth]{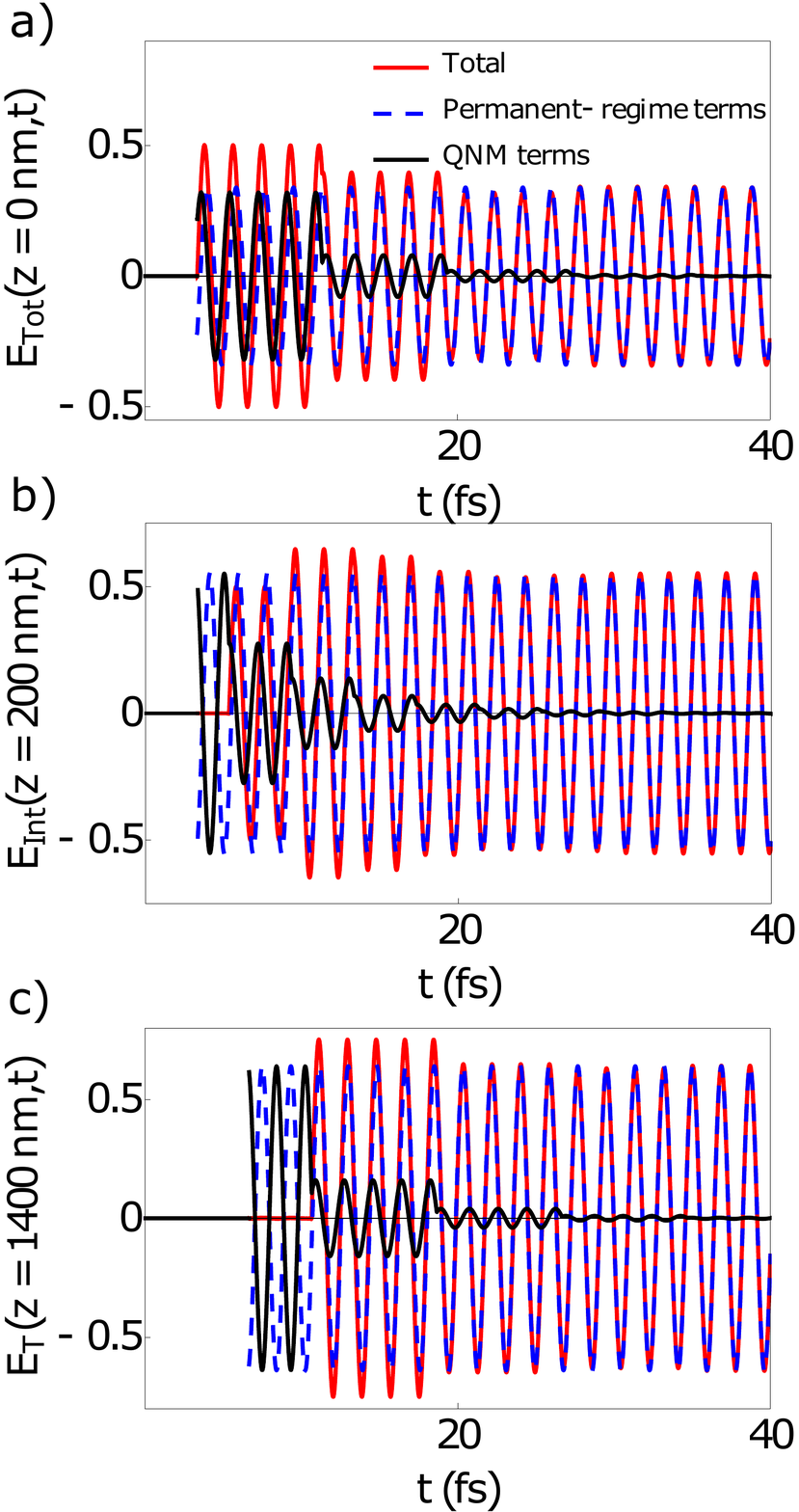}
\end{center}
\caption{Total electric field amplitude when the slab is excited by the causal sinusoidal electric field $E_{\text{inc}}(z=0,t) =H\left(t\right)\sin\left(\omega_0 t\right)$. Total (incident+reflected) electric field calculated on the incident medium ($z=0$~nm (a)), internal field calculated in middle of the slab ($z=200$~nm, (b)), and transmitted field calculated 1000~nm after the slab ($z=1400$~nm (c)). The electric field amplitude is calculated with the full expression in Eq.~\ref{E_R_express_bis} (and with Eqs.~(43,44) of the Supplemental material) (total; full red lines), with the first term that depends on $\omega_0$ in Eq.~\ref{E_R_express_bis} (permanent regime term; dashed blue lines) and the second term that depends on $\omega_{p,\alpha}^{(i)}$ (transient regime term; black full lines) in Eq.~\ref{E_R_express_bis}.}
\label{Fig:Time_dep_Field_cutoff_perm_trans_terms} 
\end{figure}

Similar expressions of the transmitted and internal fields are provided in Eqs.~(43,44) of the Supplemental Material. When comparing the results obtained with these formulas to numerical results computed with CST, a very good agreement is observed (see Fig.~(5) of the  Supplemental Material). In the expressions of the fields, there are terms that are proportional to $e^{-i \omega_{p,\alpha}^{(e)} t}$ and also to the field of the associated QNM modes. However, these terms decrease exponentially with respect to time and consequently vanish when $t\rightarrow\infty$ since $\Im\left(\omega_{p,\alpha}^{(i)}\right)<0$: the contribution of QNMs exponentially decrease with respect to time. On the other hand, terms proportional to $e^{-i \omega_{0} t}$ have a constant amplitude in time and describe the response of the resonator in the permanent regime. 

The transient response of the resonator consequently corresponds to an interference between the resonant terms and the permanent response (see Fig.~\ref{Fig:Time_dep_Field_cutoff_perm_trans_terms}). This result is consistent with the fact that linear systems illuminated with a monochromatic wave at frequency $\omega_0$ feature a monochromatic optical response at the same frequency $\omega_0$ and cannot feature terms at the QNM frequencies $\omega_n$. These terms can play a role in the transient regime since the excitation field exhibits a wide spectrum at short times. Consequently, in the harmonic regime, there are no terms proportional to the QNM fields. In this case, the eigen-modes only appear as poles of the transmission or reflection coefficients. Let us remark that the opposite result is observed when a sinusoidal excitation is shut-off: the signal was monochromatic before the cut-off and the signal after the cut-off results from the interference between the different modal contributions before fading off \cite{dubois2015time}. 

To conclude, we derived the Impulse Response Function (IRF) of optical cavities with respect to their resonant states. This work provides an important step toward the derivation of the modal expansion of the $time-dependent$ Green function. We studied the influence of resonant states on the response of optical cavities in the time domain and demonstrated that quasi-normal modes contribute to the optical response only in the transient regime. The optical response converges towards the harmonic case when $t \rightarrow \infty$ and the field does not feature any term proportional to QNMs. This result proves that the modal analysis of optical cavities in the harmonic domain requires the calculation of the pole expansion of the $S$-matrix, similarly to the transfer function in electronics.
\section{ACKNOWLEDGMENTS}
The authors thank Marc Dubois and Elena Mikheeva for their support with full numerical simulations and stimulating discussions. 


\begin{thebibliography}{29}%
\makeatletter
\providecommand \@ifxundefined [1]{%
 \@ifx{#1\undefined}
}%
\providecommand \@ifnum [1]{%
 \ifnum #1\expandafter \@firstoftwo
 \else \expandafter \@secondoftwo
 \fi
}%
\providecommand \@ifx [1]{%
 \ifx #1\expandafter \@firstoftwo
 \else \expandafter \@secondoftwo
 \fi
}%
\providecommand \natexlab [1]{#1}%
\providecommand \enquote  [1]{``#1''}%
\providecommand \bibnamefont  [1]{#1}%
\providecommand \bibfnamefont [1]{#1}%
\providecommand \citenamefont [1]{#1}%
\providecommand \href@noop [0]{\@secondoftwo}%
\providecommand \href [0]{\begingroup \@sanitize@url \@href}%
\providecommand \@href[1]{\@@startlink{#1}\@@href}%
\providecommand \@@href[1]{\endgroup#1\@@endlink}%
\providecommand \@sanitize@url [0]{\catcode `\\12\catcode `\$12\catcode
  `\&12\catcode `\#12\catcode `\^12\catcode `\_12\catcode `\%12\relax}%
\providecommand \@@startlink[1]{}%
\providecommand \@@endlink[0]{}%
\providecommand \url  [0]{\begingroup\@sanitize@url \@url }%
\providecommand \@url [1]{\endgroup\@href {#1}{\urlprefix }}%
\providecommand \urlprefix  [0]{URL }%
\providecommand \Eprint [0]{\href }%
\providecommand \doibase [0]{http://dx.doi.org/}%
\providecommand \selectlanguage [0]{\@gobble}%
\providecommand \bibinfo  [0]{\@secondoftwo}%
\providecommand \bibfield  [0]{\@secondoftwo}%
\providecommand \translation [1]{[#1]}%
\providecommand \BibitemOpen [0]{}%
\providecommand \bibitemStop [0]{}%
\providecommand \bibitemNoStop [0]{.\EOS\space}%
\providecommand \EOS [0]{\spacefactor3000\relax}%
\providecommand \BibitemShut  [1]{\csname bibitem#1\endcsname}%
\let\auto@bib@innerbib\@empty
\bibitem [{\citenamefont {Sandoghdar}\ \emph {et~al.}(1996)\citenamefont
  {Sandoghdar}, \citenamefont {Treussart}, \citenamefont {Hare}, \citenamefont
  {Lefevre-Seguin}, \citenamefont {Raimond},\ and\ \citenamefont
  {Haroche}}]{sandoghdar1996very}%
  \BibitemOpen
  \bibfield  {author} {\bibinfo {author} {\bibfnamefont {V.}~\bibnamefont
  {Sandoghdar}}, \bibinfo {author} {\bibfnamefont {F.}~\bibnamefont
  {Treussart}}, \bibinfo {author} {\bibfnamefont {J.}~\bibnamefont {Hare}},
  \bibinfo {author} {\bibfnamefont {V.}~\bibnamefont {Lefevre-Seguin}},
  \bibinfo {author} {\bibfnamefont {J.-M.}\ \bibnamefont {Raimond}}, \ and\
  \bibinfo {author} {\bibfnamefont {S.}~\bibnamefont {Haroche}},\ }\href@noop
  {} {\bibfield  {journal} {\bibinfo  {journal} {Physical review A}\ }\textbf
  {\bibinfo {volume} {54}},\ \bibinfo {pages} {R1777} (\bibinfo {year}
  {1996})}\BibitemShut {NoStop}%
\bibitem [{\citenamefont {Del’Haye}\ \emph {et~al.}(2007)\citenamefont
  {Del’Haye}, \citenamefont {Schliesser}, \citenamefont {Arcizet},
  \citenamefont {Wilken}, \citenamefont {Holzwarth},\ and\ \citenamefont
  {Kippenberg}}]{del2007optical}%
  \BibitemOpen
  \bibfield  {author} {\bibinfo {author} {\bibfnamefont {P.}~\bibnamefont
  {Del’Haye}}, \bibinfo {author} {\bibfnamefont {A.}~\bibnamefont
  {Schliesser}}, \bibinfo {author} {\bibfnamefont {O.}~\bibnamefont {Arcizet}},
  \bibinfo {author} {\bibfnamefont {T.}~\bibnamefont {Wilken}}, \bibinfo
  {author} {\bibfnamefont {R.}~\bibnamefont {Holzwarth}}, \ and\ \bibinfo
  {author} {\bibfnamefont {T.~J.}\ \bibnamefont {Kippenberg}},\ }\href@noop {}
  {\bibfield  {journal} {\bibinfo  {journal} {Nature}\ }\textbf {\bibinfo
  {volume} {450}},\ \bibinfo {pages} {1214} (\bibinfo {year}
  {2007})}\BibitemShut {NoStop}%
\bibitem [{\citenamefont {Somaschi}\ \emph {et~al.}(2016)\citenamefont
  {Somaschi}, \citenamefont {Giesz}, \citenamefont {De~Santis}, \citenamefont
  {Loredo}, \citenamefont {Almeida}, \citenamefont {Hornecker}, \citenamefont
  {Portalupi}, \citenamefont {Grange}, \citenamefont {Ant{\'o}n}, \citenamefont
  {Demory} \emph {et~al.}}]{somaschi2016near}%
  \BibitemOpen
  \bibfield  {author} {\bibinfo {author} {\bibfnamefont {N.}~\bibnamefont
  {Somaschi}}, \bibinfo {author} {\bibfnamefont {V.}~\bibnamefont {Giesz}},
  \bibinfo {author} {\bibfnamefont {L.}~\bibnamefont {De~Santis}}, \bibinfo
  {author} {\bibfnamefont {J.}~\bibnamefont {Loredo}}, \bibinfo {author}
  {\bibfnamefont {M.~P.}\ \bibnamefont {Almeida}}, \bibinfo {author}
  {\bibfnamefont {G.}~\bibnamefont {Hornecker}}, \bibinfo {author}
  {\bibfnamefont {S.~L.}\ \bibnamefont {Portalupi}}, \bibinfo {author}
  {\bibfnamefont {T.}~\bibnamefont {Grange}}, \bibinfo {author} {\bibfnamefont
  {C.}~\bibnamefont {Ant{\'o}n}}, \bibinfo {author} {\bibfnamefont
  {J.}~\bibnamefont {Demory}},  \emph {et~al.},\ }\href@noop {} {\bibfield
  {journal} {\bibinfo  {journal} {Nature Photonics}\ }\textbf {\bibinfo
  {volume} {10}},\ \bibinfo {pages} {340} (\bibinfo {year} {2016})}\BibitemShut
  {NoStop}%
\bibitem [{\citenamefont {Born}\ and\ \citenamefont {Wolf}(2000)}]{Born00}%
  \BibitemOpen
  \bibfield  {author} {\bibinfo {author} {\bibfnamefont {M.}~\bibnamefont
  {Born}}\ and\ \bibinfo {author} {\bibfnamefont {E.}~\bibnamefont {Wolf}},\
  }\href {http://books.google.fr/books?id=oV80AAAAIAAJ} {\emph {\bibinfo
  {title} {Principles of Optics: Electromagnetic Theory of Propagation,
  Interference and Diffraction of Light, seventh edition}}},\ edited by\
  \bibinfo {editor} {\bibfnamefont {C.}~\bibnamefont {Archive}}\ (\bibinfo
  {publisher} {Cambridge University Press},\ \bibinfo {year}
  {2000})\BibitemShut {NoStop}%
\bibitem [{\citenamefont {Baranov}\ \emph {et~al.}(2017)\citenamefont
  {Baranov}, \citenamefont {Krasnok},\ and\ \citenamefont
  {Al{\`u}}}]{baranov2017coherent}%
  \BibitemOpen
  \bibfield  {author} {\bibinfo {author} {\bibfnamefont {D.~G.}\ \bibnamefont
  {Baranov}}, \bibinfo {author} {\bibfnamefont {A.}~\bibnamefont {Krasnok}}, \
  and\ \bibinfo {author} {\bibfnamefont {A.}~\bibnamefont {Al{\`u}}},\
  }\href@noop {} {\bibfield  {journal} {\bibinfo  {journal} {Optica}\ }\textbf
  {\bibinfo {volume} {4}},\ \bibinfo {pages} {1457} (\bibinfo {year}
  {2017})}\BibitemShut {NoStop}%
\bibitem [{\citenamefont {Lalanne}\ \emph {et~al.}(2018)\citenamefont
  {Lalanne}, \citenamefont {Yan}, \citenamefont {Vynck}, \citenamefont
  {Sauvan},\ and\ \citenamefont {Hugonin}}]{lalanne2018light}%
  \BibitemOpen
  \bibfield  {author} {\bibinfo {author} {\bibfnamefont {P.}~\bibnamefont
  {Lalanne}}, \bibinfo {author} {\bibfnamefont {W.}~\bibnamefont {Yan}},
  \bibinfo {author} {\bibfnamefont {K.}~\bibnamefont {Vynck}}, \bibinfo
  {author} {\bibfnamefont {C.}~\bibnamefont {Sauvan}}, \ and\ \bibinfo {author}
  {\bibfnamefont {J.-P.}\ \bibnamefont {Hugonin}},\ }\href@noop {} {\bibfield
  {journal} {\bibinfo  {journal} {Laser \& Photonics Reviews}\ }\textbf
  {\bibinfo {volume} {12}},\ \bibinfo {pages} {1700113} (\bibinfo {year}
  {2018})}\BibitemShut {NoStop}%
\bibitem [{\citenamefont {Kristensen}\ \emph {et~al.}(2019)\citenamefont
  {Kristensen}, \citenamefont {Herrmann}, \citenamefont {Intravaia},\ and\
  \citenamefont {Busch}}]{Kristensen2019}%
  \BibitemOpen
  \bibfield  {author} {\bibinfo {author} {\bibfnamefont {P.~T.}\ \bibnamefont
  {Kristensen}}, \bibinfo {author} {\bibfnamefont {K.}~\bibnamefont
  {Herrmann}}, \bibinfo {author} {\bibfnamefont {F.}~\bibnamefont {Intravaia}},
  \ and\ \bibinfo {author} {\bibfnamefont {K.}~\bibnamefont {Busch}},\
  }\href@noop {} {\bibfield  {journal} {\bibinfo  {journal} {arXiv preprint
  arXiv:1910.05412}\ } (\bibinfo {year} {2019})}\BibitemShut {NoStop}%
\bibitem [{\citenamefont {Krasnok}\ \emph {et~al.}(2019)\citenamefont
  {Krasnok}, \citenamefont {Baranov}, \citenamefont {Li}, \citenamefont {Miri},
  \citenamefont {Monticone},\ and\ \citenamefont
  {Al{\'u}}}]{krasnok2019anomalies}%
  \BibitemOpen
  \bibfield  {author} {\bibinfo {author} {\bibfnamefont {A.}~\bibnamefont
  {Krasnok}}, \bibinfo {author} {\bibfnamefont {D.}~\bibnamefont {Baranov}},
  \bibinfo {author} {\bibfnamefont {H.}~\bibnamefont {Li}}, \bibinfo {author}
  {\bibfnamefont {M.-A.}\ \bibnamefont {Miri}}, \bibinfo {author}
  {\bibfnamefont {F.}~\bibnamefont {Monticone}}, \ and\ \bibinfo {author}
  {\bibfnamefont {A.}~\bibnamefont {Al{\'u}}},\ }\href@noop {} {\bibfield
  {journal} {\bibinfo  {journal} {Advances in Optics and Photonics}\ }\textbf
  {\bibinfo {volume} {11}},\ \bibinfo {pages} {892} (\bibinfo {year}
  {2019})}\BibitemShut {NoStop}%
\bibitem [{\citenamefont {Muljarov}\ \emph {et~al.}(2011)\citenamefont
  {Muljarov}, \citenamefont {Langbein},\ and\ \citenamefont
  {Zimmermann}}]{muljarov2011brillouin}%
  \BibitemOpen
  \bibfield  {author} {\bibinfo {author} {\bibfnamefont {E.}~\bibnamefont
  {Muljarov}}, \bibinfo {author} {\bibfnamefont {W.}~\bibnamefont {Langbein}},
  \ and\ \bibinfo {author} {\bibfnamefont {R.}~\bibnamefont {Zimmermann}},\
  }\href@noop {} {\bibfield  {journal} {\bibinfo  {journal} {EPL}\ }\textbf
  {\bibinfo {volume} {92}},\ \bibinfo {pages} {50010} (\bibinfo {year}
  {2011})}\BibitemShut {NoStop}%
\bibitem [{\citenamefont {Sauvan}\ \emph {et~al.}(2013)\citenamefont {Sauvan},
  \citenamefont {Hugonin}, \citenamefont {Maksymov},\ and\ \citenamefont
  {Lalanne}}]{Sauvan2013}%
  \BibitemOpen
  \bibfield  {author} {\bibinfo {author} {\bibfnamefont {C.}~\bibnamefont
  {Sauvan}}, \bibinfo {author} {\bibfnamefont {J.-P.}\ \bibnamefont {Hugonin}},
  \bibinfo {author} {\bibfnamefont {I.}~\bibnamefont {Maksymov}}, \ and\
  \bibinfo {author} {\bibfnamefont {P.}~\bibnamefont {Lalanne}},\ }\href@noop
  {} {\bibfield  {journal} {\bibinfo  {journal} {Phys. Rev. Lett.}\ }\textbf
  {\bibinfo {volume} {110}},\ \bibinfo {pages} {237401} (\bibinfo {year}
  {2013})}\BibitemShut {NoStop}%
\bibitem [{\citenamefont {Vial}\ \emph {et~al.}(2014)\citenamefont {Vial},
  \citenamefont {Zolla}, \citenamefont {Nicolet},\ and\ \citenamefont
  {Commandr{\'e}}}]{Vial2014}%
  \BibitemOpen
  \bibfield  {author} {\bibinfo {author} {\bibfnamefont {B.}~\bibnamefont
  {Vial}}, \bibinfo {author} {\bibfnamefont {F.}~\bibnamefont {Zolla}},
  \bibinfo {author} {\bibfnamefont {A.}~\bibnamefont {Nicolet}}, \ and\
  \bibinfo {author} {\bibfnamefont {M.}~\bibnamefont {Commandr{\'e}}},\
  }\href@noop {} {\bibfield  {journal} {\bibinfo  {journal} {Phys. Rev. A}\
  }\textbf {\bibinfo {volume} {89}},\ \bibinfo {pages} {023829} (\bibinfo
  {year} {2014})}\BibitemShut {NoStop}%
\bibitem [{\citenamefont {Zambrana-Puyalto}\ and\ \citenamefont
  {Bonod}(2015)}]{zambrana2015purcell}%
  \BibitemOpen
  \bibfield  {author} {\bibinfo {author} {\bibfnamefont {X.}~\bibnamefont
  {Zambrana-Puyalto}}\ and\ \bibinfo {author} {\bibfnamefont {N.}~\bibnamefont
  {Bonod}},\ }\href@noop {} {\bibfield  {journal} {\bibinfo  {journal} {Phys.
  Rev. B}\ }\textbf {\bibinfo {volume} {91}},\ \bibinfo {pages} {195422}
  (\bibinfo {year} {2015})}\BibitemShut {NoStop}%
\bibitem [{\citenamefont {Thomson}(1883)}]{Thomson1883}%
  \BibitemOpen
  \bibfield  {author} {\bibinfo {author} {\bibfnamefont {J.~J.}\ \bibnamefont
  {Thomson}},\ }\href@noop {} {\bibfield  {journal} {\bibinfo  {journal} {Proc.
  London Math. Soc.}\ }\textbf {\bibinfo {volume} {1}},\ \bibinfo {pages} {197}
  (\bibinfo {year} {1883})}\BibitemShut {NoStop}%
\bibitem [{\citenamefont {Lamb}(1900)}]{Lamb1900}%
  \BibitemOpen
  \bibfield  {author} {\bibinfo {author} {\bibfnamefont {H.}~\bibnamefont
  {Lamb}},\ }\href@noop {} {\bibfield  {journal} {\bibinfo  {journal} {Proc.
  London Math. Soc.}\ }\textbf {\bibinfo {volume} {1}},\ \bibinfo {pages} {208}
  (\bibinfo {year} {1900})}\BibitemShut {NoStop}%
\bibitem [{\citenamefont {Nussenzveig}(1972)}]{nussenzveig1972causality}%
  \BibitemOpen
  \bibfield  {author} {\bibinfo {author} {\bibfnamefont {H.~M.}\ \bibnamefont
  {Nussenzveig}},\ }\href@noop {} {\emph {\bibinfo {title} {Causality and
  dispersion relations}}}\ (\bibinfo  {publisher} {Academic Press New York},\
  \bibinfo {year} {1972})\BibitemShut {NoStop}%
\bibitem [{\citenamefont {Faggiani}\ \emph {et~al.}(2017)\citenamefont
  {Faggiani}, \citenamefont {Losquin}, \citenamefont {Yang}, \citenamefont
  {Marsell}, \citenamefont {Mikkelsen},\ and\ \citenamefont
  {Lalanne}}]{faggiani2017modal}%
  \BibitemOpen
  \bibfield  {author} {\bibinfo {author} {\bibfnamefont {R.}~\bibnamefont
  {Faggiani}}, \bibinfo {author} {\bibfnamefont {A.}~\bibnamefont {Losquin}},
  \bibinfo {author} {\bibfnamefont {J.}~\bibnamefont {Yang}}, \bibinfo {author}
  {\bibfnamefont {E.}~\bibnamefont {Marsell}}, \bibinfo {author} {\bibfnamefont
  {A.}~\bibnamefont {Mikkelsen}}, \ and\ \bibinfo {author} {\bibfnamefont
  {P.}~\bibnamefont {Lalanne}},\ }\href@noop {} {\bibfield  {journal} {\bibinfo
   {journal} {ACS Photonics}\ }\textbf {\bibinfo {volume} {4}},\ \bibinfo
  {pages} {897} (\bibinfo {year} {2017})}\BibitemShut {NoStop}%
\bibitem [{\citenamefont {Lalanne}\ \emph {et~al.}(2019)\citenamefont
  {Lalanne}, \citenamefont {Yan}, \citenamefont {Gras}, \citenamefont {Sauvan},
  \citenamefont {Hugonin}, \citenamefont {Besbes}, \citenamefont {Dem{\'e}sy},
  \citenamefont {Truong}, \citenamefont {Gralak}, \citenamefont {Zolla} \emph
  {et~al.}}]{lalanne2019quasinormal}%
  \BibitemOpen
  \bibfield  {author} {\bibinfo {author} {\bibfnamefont {P.}~\bibnamefont
  {Lalanne}}, \bibinfo {author} {\bibfnamefont {W.}~\bibnamefont {Yan}},
  \bibinfo {author} {\bibfnamefont {A.}~\bibnamefont {Gras}}, \bibinfo {author}
  {\bibfnamefont {C.}~\bibnamefont {Sauvan}}, \bibinfo {author} {\bibfnamefont
  {J.-P.}\ \bibnamefont {Hugonin}}, \bibinfo {author} {\bibfnamefont
  {M.}~\bibnamefont {Besbes}}, \bibinfo {author} {\bibfnamefont
  {G.}~\bibnamefont {Dem{\'e}sy}}, \bibinfo {author} {\bibfnamefont
  {M.}~\bibnamefont {Truong}}, \bibinfo {author} {\bibfnamefont
  {B.}~\bibnamefont {Gralak}}, \bibinfo {author} {\bibfnamefont
  {F.}~\bibnamefont {Zolla}},  \emph {et~al.},\ }\href@noop {} {\bibfield
  {journal} {\bibinfo  {journal} {JOSA A}\ }\textbf {\bibinfo {volume} {36}},\
  \bibinfo {pages} {686} (\bibinfo {year} {2019})}\BibitemShut {NoStop}%
\bibitem [{\citenamefont {Yan}\ \emph {et~al.}(2018)\citenamefont {Yan},
  \citenamefont {Faggiani},\ and\ \citenamefont {Lalanne}}]{Yan2018}%
  \BibitemOpen
  \bibfield  {author} {\bibinfo {author} {\bibfnamefont {W.}~\bibnamefont
  {Yan}}, \bibinfo {author} {\bibfnamefont {R.}~\bibnamefont {Faggiani}}, \
  and\ \bibinfo {author} {\bibfnamefont {P.}~\bibnamefont {Lalanne}},\
  }\href@noop {} {\bibfield  {journal} {\bibinfo  {journal} {Physical Review
  B}\ }\textbf {\bibinfo {volume} {97}},\ \bibinfo {pages} {205422} (\bibinfo
  {year} {2018})}\BibitemShut {NoStop}%
\bibitem [{\citenamefont {Stout}\ \emph {et~al.}(2019)\citenamefont {Stout},
  \citenamefont {Colom}, \citenamefont {Bonod},\ and\ \citenamefont
  {McPhedran}}]{stout2019eigenstate}%
  \BibitemOpen
  \bibfield  {author} {\bibinfo {author} {\bibfnamefont {B.}~\bibnamefont
  {Stout}}, \bibinfo {author} {\bibfnamefont {R.}~\bibnamefont {Colom}},
  \bibinfo {author} {\bibfnamefont {N.}~\bibnamefont {Bonod}}, \ and\ \bibinfo
  {author} {\bibfnamefont {R.}~\bibnamefont {McPhedran}},\ }\href@noop {}
  {\bibfield  {journal} {\bibinfo  {journal} {arXiv preprint arXiv:1903.07183}\
  } (\bibinfo {year} {2019})}\BibitemShut {NoStop}%
\bibitem [{\citenamefont {Colom}\ \emph {et~al.}(2018)\citenamefont {Colom},
  \citenamefont {McPhedran}, \citenamefont {Stout},\ and\ \citenamefont
  {Bonod}}]{Colom2018}%
  \BibitemOpen
  \bibfield  {author} {\bibinfo {author} {\bibfnamefont {R.}~\bibnamefont
  {Colom}}, \bibinfo {author} {\bibfnamefont {R.~C.}\ \bibnamefont
  {McPhedran}}, \bibinfo {author} {\bibfnamefont {B.}~\bibnamefont {Stout}}, \
  and\ \bibinfo {author} {\bibfnamefont {N.}~\bibnamefont {Bonod}},\
  }\href@noop {} {\bibfield  {journal} {\bibinfo  {journal} {Phys. Rev. B}\
  }\textbf {\bibinfo {volume} {98}},\ \bibinfo {pages} {005400} (\bibinfo
  {year} {2018})}\BibitemShut {NoStop}%
\bibitem [{\citenamefont {Abdelrahman}\ and\ \citenamefont
  {Gralak}(2018)}]{abdelrahman2018completeness}%
  \BibitemOpen
  \bibfield  {author} {\bibinfo {author} {\bibfnamefont {M.~I.}\ \bibnamefont
  {Abdelrahman}}\ and\ \bibinfo {author} {\bibfnamefont {B.}~\bibnamefont
  {Gralak}},\ }\href@noop {} {\bibfield  {journal} {\bibinfo  {journal} {OSA
  Continuum}\ }\textbf {\bibinfo {volume} {1}},\ \bibinfo {pages} {340}
  (\bibinfo {year} {2018})}\BibitemShut {NoStop}%
\bibitem [{\citenamefont {Zschiedrich}\ \emph {et~al.}(2018)\citenamefont
  {Zschiedrich}, \citenamefont {Binkowski}, \citenamefont {Nikolay},
  \citenamefont {Benson}, \citenamefont {Kewes},\ and\ \citenamefont
  {Burger}}]{Zschiedrich2018}%
  \BibitemOpen
  \bibfield  {author} {\bibinfo {author} {\bibfnamefont {L.}~\bibnamefont
  {Zschiedrich}}, \bibinfo {author} {\bibfnamefont {F.}~\bibnamefont
  {Binkowski}}, \bibinfo {author} {\bibfnamefont {N.}~\bibnamefont {Nikolay}},
  \bibinfo {author} {\bibfnamefont {O.}~\bibnamefont {Benson}}, \bibinfo
  {author} {\bibfnamefont {G.}~\bibnamefont {Kewes}}, \ and\ \bibinfo {author}
  {\bibfnamefont {S.}~\bibnamefont {Burger}},\ }\href@noop {} {\bibfield
  {journal} {\bibinfo  {journal} {Physical Review A}\ }\textbf {\bibinfo
  {volume} {98}},\ \bibinfo {pages} {043806} (\bibinfo {year}
  {2018})}\BibitemShut {NoStop}%
\bibitem [{\citenamefont {Leung}\ and\ \citenamefont {Pang}(1996)}]{Leung1996}%
  \BibitemOpen
  \bibfield  {author} {\bibinfo {author} {\bibfnamefont {P.}~\bibnamefont
  {Leung}}\ and\ \bibinfo {author} {\bibfnamefont {K.}~\bibnamefont {Pang}},\
  }\href@noop {} {\bibfield  {journal} {\bibinfo  {journal} {J. Opt. Soc. Am.
  B}\ }\textbf {\bibinfo {volume} {13}},\ \bibinfo {pages} {805} (\bibinfo
  {year} {1996})}\BibitemShut {NoStop}%
\bibitem [{\citenamefont {Gippius}\ \emph {et~al.}(2010)\citenamefont
  {Gippius}, \citenamefont {Weiss}, \citenamefont {Tikhodeev},\ and\
  \citenamefont {Giessen}}]{Gippius2010}%
  \BibitemOpen
  \bibfield  {author} {\bibinfo {author} {\bibfnamefont {N.~A.}\ \bibnamefont
  {Gippius}}, \bibinfo {author} {\bibfnamefont {T.}~\bibnamefont {Weiss}},
  \bibinfo {author} {\bibfnamefont {S.~G.}\ \bibnamefont {Tikhodeev}}, \ and\
  \bibinfo {author} {\bibfnamefont {H.}~\bibnamefont {Giessen}},\ }\href@noop
  {} {\bibfield  {journal} {\bibinfo  {journal} {Optics Express}\ }\textbf
  {\bibinfo {volume} {18}},\ \bibinfo {pages} {7569} (\bibinfo {year}
  {2010})}\BibitemShut {NoStop}%
\bibitem [{\citenamefont {Grigoriev}\ \emph {et~al.}(2013)\citenamefont
  {Grigoriev}, \citenamefont {Varault}, \citenamefont {Boudarham},
  \citenamefont {Stout}, \citenamefont {Wenger},\ and\ \citenamefont
  {Bonod}}]{grigoriev2013singular}%
  \BibitemOpen
  \bibfield  {author} {\bibinfo {author} {\bibfnamefont {V.}~\bibnamefont
  {Grigoriev}}, \bibinfo {author} {\bibfnamefont {S.}~\bibnamefont {Varault}},
  \bibinfo {author} {\bibfnamefont {G.}~\bibnamefont {Boudarham}}, \bibinfo
  {author} {\bibfnamefont {B.}~\bibnamefont {Stout}}, \bibinfo {author}
  {\bibfnamefont {J.}~\bibnamefont {Wenger}}, \ and\ \bibinfo {author}
  {\bibfnamefont {N.}~\bibnamefont {Bonod}},\ }\href@noop {} {\bibfield
  {journal} {\bibinfo  {journal} {Physical Review A}\ }\textbf {\bibinfo
  {volume} {88}},\ \bibinfo {pages} {063805} (\bibinfo {year}
  {2013})}\BibitemShut {NoStop}%
\bibitem [{\citenamefont {Weiss}\ and\ \citenamefont
  {Muljarov}(2018)}]{Weiss2018}%
  \BibitemOpen
  \bibfield  {author} {\bibinfo {author} {\bibfnamefont {T.}~\bibnamefont
  {Weiss}}\ and\ \bibinfo {author} {\bibfnamefont {E.}~\bibnamefont
  {Muljarov}},\ }\href@noop {} {\bibfield  {journal} {\bibinfo  {journal}
  {Physical Review B}\ }\textbf {\bibinfo {volume} {98}},\ \bibinfo {pages}
  {085433} (\bibinfo {year} {2018})}\BibitemShut {NoStop}%
\bibitem [{\citenamefont {Dem{\'e}sy}\ \emph {et~al.}(2018)\citenamefont
  {Dem{\'e}sy}, \citenamefont {Auger},\ and\ \citenamefont
  {Stout}}]{demesy2018scattering}%
  \BibitemOpen
  \bibfield  {author} {\bibinfo {author} {\bibfnamefont {G.}~\bibnamefont
  {Dem{\'e}sy}}, \bibinfo {author} {\bibfnamefont {J.-C.}\ \bibnamefont
  {Auger}}, \ and\ \bibinfo {author} {\bibfnamefont {B.}~\bibnamefont
  {Stout}},\ }\href@noop {} {\bibfield  {journal} {\bibinfo  {journal} {JOSA
  A}\ }\textbf {\bibinfo {volume} {35}},\ \bibinfo {pages} {1401} (\bibinfo
  {year} {2018})}\BibitemShut {NoStop}%
\bibitem [{\citenamefont {Colom}\ \emph {et~al.}(2019)\citenamefont {Colom},
  \citenamefont {McPhedran}, \citenamefont {Stout},\ and\ \citenamefont
  {Bonod}}]{colom2019modal}%
  \BibitemOpen
  \bibfield  {author} {\bibinfo {author} {\bibfnamefont {R.}~\bibnamefont
  {Colom}}, \bibinfo {author} {\bibfnamefont {R.}~\bibnamefont {McPhedran}},
  \bibinfo {author} {\bibfnamefont {B.}~\bibnamefont {Stout}}, \ and\ \bibinfo
  {author} {\bibfnamefont {N.}~\bibnamefont {Bonod}},\ }\href@noop {}
  {\bibfield  {journal} {\bibinfo  {journal} {JOSA B}\ }\textbf {\bibinfo
  {volume} {36}},\ \bibinfo {pages} {2052} (\bibinfo {year}
  {2019})}\BibitemShut {NoStop}%
\bibitem [{\citenamefont {Dubois}\ \emph {et~al.}(2015)\citenamefont {Dubois},
  \citenamefont {Bossy}, \citenamefont {Enoch}, \citenamefont {Guenneau},
  \citenamefont {Lerosey},\ and\ \citenamefont {Sebbah}}]{dubois2015time}%
  \BibitemOpen
  \bibfield  {author} {\bibinfo {author} {\bibfnamefont {M.}~\bibnamefont
  {Dubois}}, \bibinfo {author} {\bibfnamefont {E.}~\bibnamefont {Bossy}},
  \bibinfo {author} {\bibfnamefont {S.}~\bibnamefont {Enoch}}, \bibinfo
  {author} {\bibfnamefont {S.}~\bibnamefont {Guenneau}}, \bibinfo {author}
  {\bibfnamefont {G.}~\bibnamefont {Lerosey}}, \ and\ \bibinfo {author}
  {\bibfnamefont {P.}~\bibnamefont {Sebbah}},\ }\href@noop {} {\bibfield
  {journal} {\bibinfo  {journal} {Physical review letters}\ }\textbf {\bibinfo
  {volume} {114}},\ \bibinfo {pages} {013902} (\bibinfo {year}
  {2015})}\BibitemShut {NoStop}%
\end{thebibliography}%
%

\end{document}